\documentclass[showpacs,aps,pre,twocolumn,floatfix]{revtex4}
\usepackage{graphicx}

\begin{document}

\title{Interaction of streamers and stationary corrugated ionization waves in semiconductors}
\author{A. S. Kyuregyan}\email[E-mail: ]{ask@vei.ru}
\affiliation{All-Russia Electrical Engineering Institute, 12 Krasnokazarmennaya St., 111250 Moscow, Russian Federation}

\begin{abstract}
A numerical simulation of evolution of an identical interacting streamers  array in semiconductors has been performed using the diffusion-drift approximation and taking into account the impact and tunnel ionization. It has been assumed that the external electric field $E_0$ is static and uniform, the background electrons and holes are absent, the initial avalanches start simultaneously from the nodes of the plane hexagonal lattice, which is perpendicular to the external field, however the avalanches and streamers are axially symmetric within a cylinder of radius $R$. It has been shown that under certain conditions, the interaction between the streamers leads finally either to the formation of two types of stationary ionization waves with corrugated front or to a stationary plane ionization wave. A diagram of different steady states of this type waves in the  plane of parameter $E_0,R$ has been presented and a qualitative explanation of the plane partition into four different regions has been given. Characteristics of corrugated waves have been studied in detail and discussed in the region of $R$ and $E_0$ large values, in which the maximum field strength at the front is large enough for the tunnel ionization implementation. It has been shown that corrugated waves ionize semiconductor more efficiently than flat ones, especially in relatively weak external fields.
\end{abstract}


\maketitle

\section{Introduction}

The streamer mechanism of electric discharge is used for a long time for the description of pulse breakdown of various matters. A set of works is devoted to theoretical and experimental studying of streamers, however between these two ways of research there is an essential discrepancy. In vast majority of the theoretical works performed both by analytical and numerical methods, single streamers were studied. Meanwhile in practice the discharge is usually carried out by a large number of interacting  streamers. Such electrostatic interaction has to be essential, in particular, in a pulse crown and in a streamer zone of a long spark where the characteristic distance between streamers is less than their lengths, but there is more than their diameters \cite{Kor,Baz}. The excellent photos especially visually illustrating this circumstance, were received for the last years (see, for example, \cite{ Nijd10,Bri08}). Modeling of such multistreamer discharges is a very complex three-dimensional problem, so as a first step towards its solution a very simplified situation should be studied  - the evolution of  an identical streamers, which simultaneously starts from nodes of one-dimensional  (in this case, the model corresponds to experiments \cite {Kras,Ono07} ) or two-dimensional periodic lattice.

As far as we know, the first attempt of this kind was made in the work \cite{Nai} whose author studied evolution of the one-dimensional periodic array of the cylindrical streamers propagating in air from a thin wire to the  plane parallel to it. Interaction between the charged  streamers was taken into account by the approximate analytical solution of the corresponding electrostatic problem  and  introduction of so obtained amendments to the field strength  in a numerical model of a single streamer. It was shown that field strength before the front of each streamer in the array and the speed of their propagation considerably decrease in comparison with a single streamer. This result is quite expected. However, the author \cite {Nai} didn't manage to receive any additional information, because in the framework of so-called 1.5D-dimensional numerical model which he used, the cross sizes and a form of each streamer are a priori set and therefore aren't subject to interstreamer interaction.

This shortage is devoided in the work \cite{Luq1}, whose  authors simulated the evolution of two-dimensional periodic array of negative streamers in gases in a uniform external field $ E_0 $, using "minimal model" (taking into account the drift, diffusion and impact ionization of electrons, background electrons are absent, additional mechanisms of ionization are not taken into account \cite{Ebe1}). At the initial stage, streamers develop independently from each other so that their length and front  curvature radius  increase with constant velocities \cite{Brau,ASK6}. But the nature of further evolution depends strongly on the distance $ 2L $  between  streamers. If $ L $ is greater than some critical value $ L_c(E_0) $,  the front of each streamer starts quickly becoming distorted as a result of transverse instability, described in \cite{Arr04,Arr05,Arr08,Derks} for gases and in \cite{ASK6,ASK8} for semiconductors. However, at $ L <L_c (E_0) $ this instability is suppressed and eventually propagation of  streamers array  becomes stable and self-similar:  all of them travel with constant velocity $ u_f $ and unaltered form of the front, which does not depend on the initial conditions and can be described by a multi-valued function
\begin{equation}\label{eq1-1}
y_f(x)=\frac{2}{\pi}bL\arccos\left[\exp\left(\frac{\pi}{2}\frac{x-x_f}{aL}\right)\right],
\end{equation}
where $ x_f = u_f t $ is a front position on the axis $ x $ parallel to the external field. This formula was obtained in \cite {Saf-Tay} to describe the shape of the interface between two incompressible fluids with very different viscosities (for example, water, which forces the glycerin)  moving  in the narrow gap between two plates with  width $ 2L $ (Hele - Shaw cell) \cite{Saff,Bens}. In this case the boundary velocity $ u_f $ in $ (1+ a / b) $ times more than velocity  $ u_0 $ of a viscous fluid far ahead the border and therefore the matter conservation law provides the relation
\begin{equation}\label{eq1-2}
a+b=1.
\end{equation}

The formula (\ref{eq1-1}) is also applicable to describe a number of other physically different but mathematically equivalent processes, provided that the normal velocity of the interface $u_n $ at each point is proportional to the gradient of some potential function $\varphi (x, y)$, which satisfies the  Laplace equation, and the interface itself is equipotential:
\begin{equation}\label{eq1-3}
\varphi(x,y_f)=const, \qquad u_n\propto E_n\equiv |\nabla_n \varphi(x,y_f)|.
\end{equation}
In experiments with Hele-Shaw cells (in this case, $ \varphi $ is pressure) it is always get $ b = 1/2 $, that is, after the establishment of the steady state motion of the interface less viscous liquid supplants  exactly half of the cell width. This nontrivial selection phenomena was explained by the influence of small, but finite deviation from the first condition in (\ref{eq1-3}) because of the surface tension \cite{Saff,Bens} or kinetic undercooling \cite{Chapman}.

Model of streamers satisfying to conditions (\ref{eq1-3}) (in this case, $ \varphi $ is electric potential), was first used in the work \cite{LozFir} (see also the book \cite{Loz3}), but in fact they can be performed only very approximately. Violation of the former is due to the finite conductivity of the plasma behind the front and nonzero thickness $\delta $ of the front, and the latter due to a more complicated dependence of $ u_n $ on the field strength $ E_n $  normal to front. Within the framework of the "minimal model" of streamers in gases, with some reservations, it is possible to use the formula \cite{Ebe1,Lag}
\begin{equation}\label{eq1-4}
u_n=u^*\equiv v_e + 2 D_e \lambda^*,
\end{equation}
where $\lambda^* = \sqrt{v_e\alpha_e/D_e}$ is a parameter of exponential decay of the electron concentration   $ n $ ahead of the front,  $ v_e $, $ D_e $ and $ \alpha_e $ are a drift velocity, a diffusion coefficient and an impact ionization coefficient, which locally depend on $ E_n $. Usually in gases second term in the right side of (\ref{eq1-4}) is relatively small \cite{Ebe1, Lag}, so at a constant electron mobility $ \mu_e $ front velocity $ u_n \approx \mu_e E_n $. This is a fact the authors of \cite{Luq1} deemed sufficient basis for applicability of the formula (\ref{eq1-1}) to interpret their results of numerical simulations, in particular, the approximate ratio \footnote{ It is a good idea for those who study performed in \cite{Luq1} array of streamers at the highest values of $ L <L_c $, when $\delta\ll L $ and the deviation from the first condition (\ref{eq1-2}) is minimal.} $ E_M \equiv E_n (x_f, 0 ) = 2E_0 $; if this is true, it was ignored that in the case of streamers there was no reason for equality (\ref{eq1-2}).

The study of a similar problem with regard to semiconductors, is also very important. In practical terms, this is due to the fact that multistreamer breakdown mechanism determines (at least in some regimes) \cite{Vain, ASK1} the operation of avalanche voltage sharpeners, which are commutators with unique characteristics \cite{Grekh,Foc}, but in the scientific  terms this is due to features  microscopic processes in semiconductors. Among them there are particularly important ones

- saturation of dependencies  $v_{e,h}(E)$ in relatively very low fields $E\sim E_s\sim 10$ kV/cm, which, in particular, leads to a significant (relative to gas), increasing the ratio $ u_n / v_ {e, h} $ even within the framework of the "minimal model" of streamers \cite{ASK3,ASK2} and

- existence of tunneling ionization, which can also increase the ratio of $ u_n / v_ {e, h} $ by orders of magnitude \cite{Rodin,R-G-05,R-R-G-05,G-R-11,ASK6,ASK7,ASK10}.

These features lead to the fact that $ u_n \propto E_n $ only at $ E_n \ll E_s $, but in the usual range of $ E_n\sim 10-10^3$ kV/cm function $ u_n (E_n) $ varies from constant to exponentially strong. Therefore, the second of the conditions (\ref{eq1-3}) is never executed, and the equation (\ref{eq1-2}) can be satisfied only by chance at some relations between the parameters of the problem. Such a radical change in the classical formulation of the problem \cite{Saff, Bens} can lead to substantially different scenarios of evolution of the interface between the phases in a highly non-equilibrium conditions.

In this paper, we tried to study this interesting problem by numerical simulation of the interaction between the streamers in semiconductors. The main aim of this work is to obtain maximally detailed complete results for the simplest case of a static uniform external field $E_0$, which can provide the basis for further research.

\section{Mathematical model of an array of streamers}

Similar to the most works on the numerical simulation of streamers, the diffusion-drift approximation is used. In this approximation, the distributions of electrons, $n(t,{\rm {\bf r}})$, and holes, $p\,(t,{\bf r})$, are described by the continuity equations, which we conveniently wrote in the form
\begin{eqnarray}
\frac{\partial(p+n)}{\partial t}+\nabla\mkern -4mu\cdot\mkern -3mu({\bf j}_h+{\bf j}_e)=2(s_g+ s_r),\label{eq2-1}\\
\frac{\partial(p-n)}{\partial t}+\nabla\mkern -4mu\cdot\mkern -3mu({\bf j}_h-{\bf j}_e )=0, \qquad\label{eq2-2}
\end{eqnarray}
where the terms $ s_ {g, r} $ describe all possible mechanisms of generation and recombination, and the free carrier flux densities are given by the expression
\begin{eqnarray}
\nonumber
{\bf j}_{e} = {\bf v}_{e} n -\nabla \left(D_{e}n\right), \quad {\bf j}_{h} = {\bf v}_{h} p -\nabla \left(D_{h}p\right),
\end{eqnarray}
where the subscripts  "$e$" and "$h$" correspond to electrons and holes, respectively. In this paper, we considered only indirect-gap semiconductors such as Si, Ge or SiC, in which the rate of radiative recombination (and, hence, photoionization rate) is very small. Therefore the generation of pairs occurs primarily due to the impact and tunnel ionization; consequently, the generation rate has the form
\begin{eqnarray}\nonumber
s_{g} = (\alpha _e v_e n + \alpha _h v_h p)h(n+p-n_{th}) + g_t ,
\end{eqnarray}
where $h(x)$ is the Heaviside unit step function,  $n_{th}$ is a certain threshold density. It is introduced in order to exclude the appearance of nonphysical solutions due to the impact ionization far ahead of the streamer front, where the densities of the electrons and holes generated by the tunnel ionization in the external field are very low
($n+p<n_{th}$) and the continual approximation is certainly inapplicable \cite{Rodin, Li1}. Under the assumption that the lifetime of charge carriers is much larger than the propagation time of the streamers, recombination is
disregarded; i.e., $s_{r}=0$ is set. Moreover, the impact ionization coefficient $\alpha _{e,h}$,  drift velocities ${\bf v}_{e,h} $, and tunnel ionization rate $g_t$ are assumed to be specified
instantaneous and local functions of the field strength $ {\bf E} (t, {\bf r}) $, satisfying the Poisson equation
\begin{equation}
\nabla \mkern -4mu\cdot\mkern -3mu{\bf E}= \frac{q}{\varepsilon}(p-n)=-\triangle\varphi,\label{eq2-3}
\end{equation}
where, $\varphi$ is electric potential, $q$ is the elementary charge\,, $\varepsilon $ is the permittivity of the semiconductor. The usual approximations,
\begin{eqnarray}
{\rm{\bf v}}_h =-{\rm{\bf v}}_e =\mu{\bf E},\quad \mu={v_s}/(E + E_s),\nonumber\\
\alpha_e=\alpha_h=\tilde{\alpha}\exp(-\tilde{E}/E),\qquad \nonumber\\
g_t = \tilde {g}_t(E/E_t)^2\exp(-E_t/E),\quad \nonumber
\end{eqnarray}
are used, and the dependence of the diffusion coefficients $D_{e,h}$ on $E$ is disregarded, i.e., $D_e = D_h = D = const$. The parameters $v_s,E_s,\tilde{\alpha},\tilde{E},\tilde {g}$ and $E_t$ are determined by the band structure of
semiconductors and electron and hole scattering mechanisms. We used the same typical values $v_s=10^{7}$cm/c, $E_s=15$~kV/cm, $\tilde{\alpha}=10^6$cm$^{-1}$, $\tilde{E}=1,5$~MV/cm, $\tilde {g}=6,7\cdot 10^{35}$cm$^{-3}$s$^{-1}$, $E_t=22,5$~MV/cm, $D=20$~cm$^{2}$c$^{-1}$ and $\varepsilon=11,8\varepsilon_0$, as in the \cite{ASK6}.

The initial conditions for the system of equations (\ref{eq2-1}) - (\ref{eq2-3}) have the form
\begin{eqnarray}
\varphi\left({0,{\bf r}}\right)=-E_0 x,\label{eq7}\quad\qquad\qquad\label{eq2-4}\\
\sigma\left({0,{\bf r}}\right)=\sigma_0\left({\bf r}-{\bf r}_i\right),\quad\rho\left(0,{\bf r}\right)=0,\label{eq2-5}
\end{eqnarray}
where $ E_0 $ is the strength of the external field, which is directed along $x$-axis,  $\sigma=q(p+n)/\varepsilon \tilde{\alpha}\tilde{E}$, $\rho=q(p-n)/\varepsilon\tilde{\alpha}\tilde{E}$ are dimensionless concentration and the space charge density of electrons and holes,  $\sigma _0\left({\bf r}\right)$ is any quite strongly localized function satisfying the normalization condition $\int{\sigma_0\left({\bf r}\right)d{\bf r}}=2q/\varepsilon\tilde{\alpha}\tilde{E}$. The Gaussian distribution is used,
\begin{equation}\nonumber
\sigma_0\left({\bf r}\right)=\sigma_0^0\exp\left({-r^2/r_\sigma^2}\right),
\end{equation}
where $\sigma_0^0=2 q/\pi^{3/2}\varepsilon\tilde{\alpha}\tilde{E}r_\sigma^3 $.
These initial conditions correspond to the appearance of one electron–hole pair at each point $ {\bf r} = {\bf r} _i $ at time $ t = 0 $. In accordance with what was said in the introduction, these points coincide with the nodes of a planar, for example, hexagonal, lattice located in the plane $ x = 0 $. In this case, avalanches and generated by them streamers have the symmetry of regular hexagonal prism. Therefore, in the cylindrical coordinate system $ {\bf r} = \left\{x, y, \vartheta \right\} $ (hereinafter $ y $ is distance from the point $ {\bf r}$ to  axis $x $, and $ \vartheta $ is azimuthal angle), strictly speaking, one should take into account the dependence of $ n, p $ and $ \varphi $ on  $ \vartheta $, that is necessary to solve the three-dimensional Cauchy problem with natural boundary conditions on the lateral faces of the prism, which requires very large computing resources. Meanwhile, the problem can be considerably simplified if  a prism with the width of the lateral faces $ H $ is approximate by a cylinder with a radius $ R = H \sqrt{ 3 \sqrt{3} / 2 \pi} \approx 0.91 H $. In this case area of the base of  prism and  cylinder are the same, and the distance between their lateral surfaces does not exceed $ 0.1H $. Such a small difference between forms  of the side surfaces  should not have a significant impact on the processes of ionization and transport near the axis of symmetry, which mainly determine the evolution of an array of streamers. To confirm the validity of this statement the main parameters (the maximum field strength on the front $ E_M $, the concentration of electrons and holes in the $ x $ axis behind the front $ \sigma^- $, and the front velocity $ u_f $) of plane,  axially and hexagonal symmetrical streamers obtained by modeling with the same finite element meshes are shown in the Table. As can be seen, the parameters of the plane streamer differ significantly from almost matching parameters of axially and hexagonal symmetrical  streamers. At the same time, the dimensionless computing speed $ \Delta x_f / \Delta t v_s $ (here $ \Delta t $ is the time spent on modeling the process of promoting the front on the distance $ \Delta x_f $) for a hexagonal streamers almost 100 times less than  for an axially symmetric one. Therefore, in this paper we neglect the dependence of $ n, p, \varphi (\vartheta) $, weak at actual values of $ y \lesssim H / 2 \approx R / 2 $, and assume that avalanches and streamers have axial symmetry.

\begin{center}
\begin{tabular}{|l|l|l|l|}
\multicolumn{4}{c}{\textbf{Parameters of various streamers at}}\\
\multicolumn{4}{c}{$E_0=0.5\tilde{E}$, $L=R=0.91H$,  $H=289/\tilde{\alpha}$}\\
\hline
\small{Parameter}&\small{Plane}&\small{Axial}&\small{Hexagonal}\\
\hline
$E_M/\tilde{E}$ & 0.773 & 1.17 & 1.18\\
\hline
$\sigma^-$ & 0.17 & 0.58 & 0.59\\
\hline
$u_f/v_s$ & 1.88 & 4.14 & 4.12\\
\hline
$\Delta x_f/\Delta t v_s$ & 0.36 & 0.15 & 0.0017 \\
\hline
\end{tabular}
\end{center}

Under the above assumptions, our task is also symmetrical with respect to the plane $ x = 0 $, so it is enough to solve it in a rectangular area
\begin{equation}
\nonumber
0 \le x \le X,\quad  0 \le y \le R,
\end{equation}
the length of which $ X $ should be much longer than streamer length. In this case, the boundary conditions take the form
\begin{eqnarray}
\nonumber
\varphi\left({t,0,y}\right)=0,\quad\left.{{\partial\varphi\left({t,x,y}\right)}/{\partial x}} \right|_{x=X}=E_0,\quad\\
\left.{{\partial\varphi\left({t,x,y}\right)}/{\partial y}} \right|_{y = 0,R } = 0,\label{eq2-6}\qquad\qquad\\
\nonumber
\sigma\left({t,X,y}\right)=0,\qquad\qquad\qquad\quad\\
\left.{{\partial\sigma\left({t,x,y}\right)}/{\partial x}}\right|_{x =0}
=\left.{{\partial\sigma\left({t,x,y}\right)}/{\partial y}} \right|_{y=0,R}=0,\label{eq2-76}\\
\nonumber
\rho\left({t,X,y}\right)=\rho\left({t,0,y}\right)=0\qquad\qquad\\
\left.{{\partial\rho\left({t,x,y}\right)}/{\partial y}}\right|_{y=0,R}=0.\label{eq2-8}\qquad\qquad
\end{eqnarray}

The Cauchy problem (\ref{eq2-1}) - (\ref{eq2-8}) was solved by the finite element method with adaptive non-uniform mesh by the way described in \cite{ASK6}.

\section{Results and discussion}

\begin{figure}[!t]\center
\includegraphics[width=230pt]{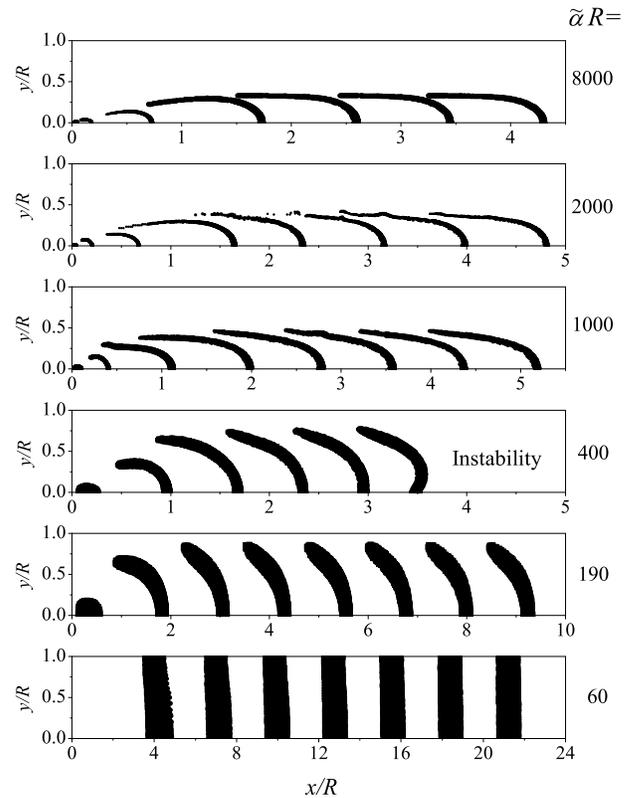}\\
\caption{The evolution of fronts of identical streamers which starts simultaneously from a planar hexagonal lattice nodes, with $ E_0 = 0.36 \tilde{E} $ and different $ R $. Black color indicates the region of the front, inside which the space charge density is greater than $0.002\varepsilon\tilde{\alpha} \tilde{E} \approx 3,2$~mC/ñm$^3$ at times  $t=i/\tilde{\alpha} v_s\approx 10i$~ps, $i=1,2...8.$} \label{fig1}
\end{figure}

\begin{figure}[]
\includegraphics[width=242pt]{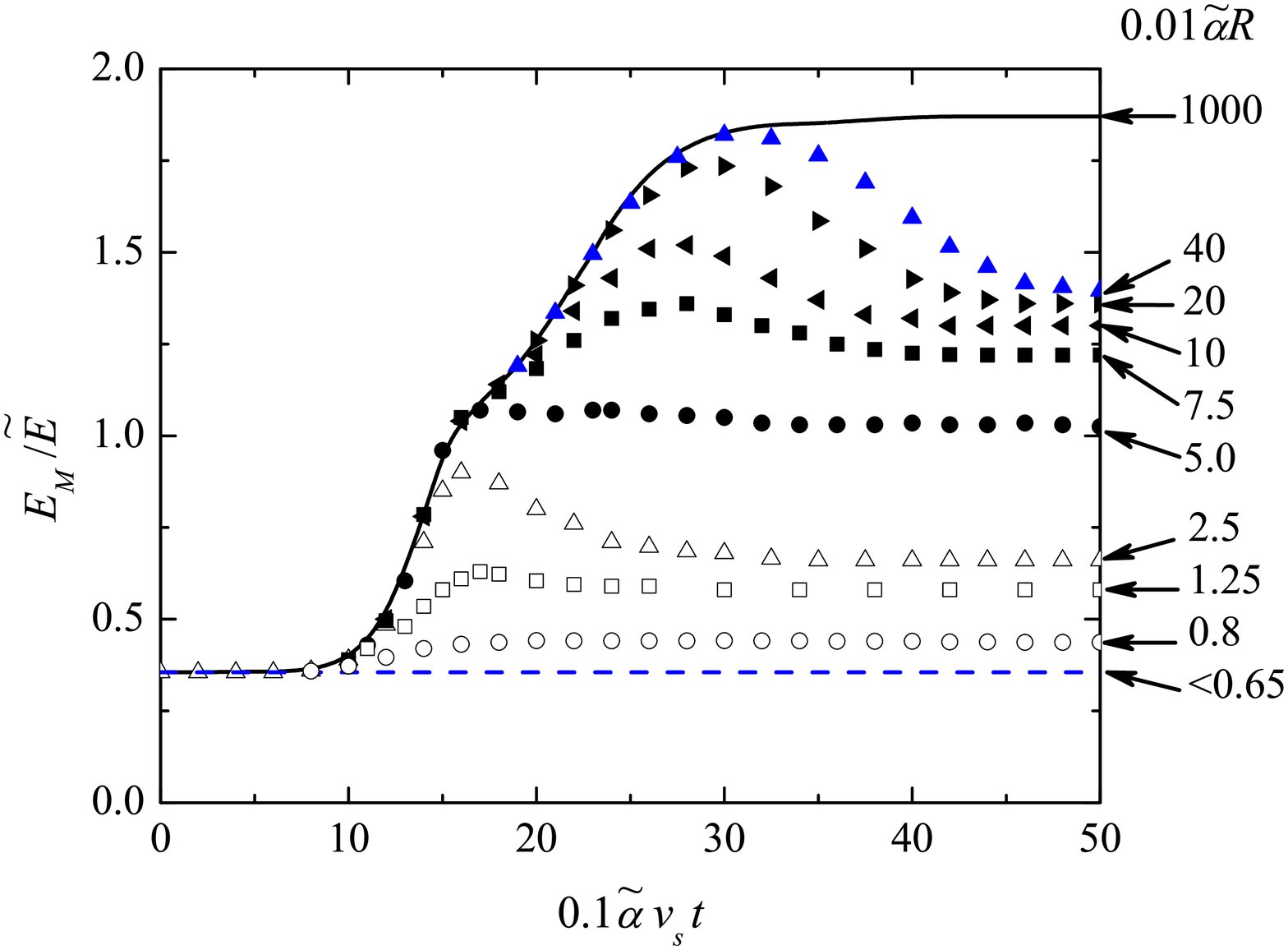}\\
\caption{Maximum field strength $ E_M $ in the streamers array vs. time at $E_0=0.36\tilde{E}$ and different values of $ R $. Open symbols: calculation with $ R_0 <R <R_c$, dark symbols: calculation with $R> R_t $, solid line: calculation for an isolated streamer,  dashed line:  the external field $E_0$.}\label{fig2}
\end{figure}

\begin{figure}[!t]\center
\includegraphics[width=230pt]{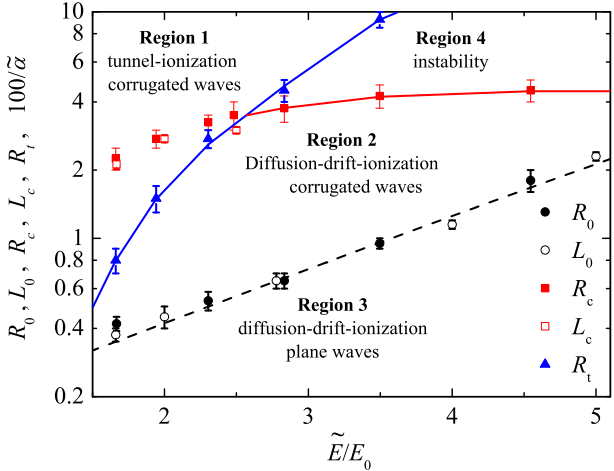}\\
\caption{Diagram of steady states of the hexagonal lattice streamers. Lines separate the plane $ (1/E_0, R) $ into four qualitatively different regions (see text). Solid squares:  calculation without tunneling ionization (with $ E_0> 0.4 \tilde{E} $), open circles: calculation for the  plane streamers array with period $ L $, open squares: the calculation for gases according to \cite{Luq1}), dashed line: calculation formulas (\ref{eq3-11}),(\ref{eq3-12}).} \label{fig3}
\end{figure}

\begin{figure}[]\center
\includegraphics[width=230pt]{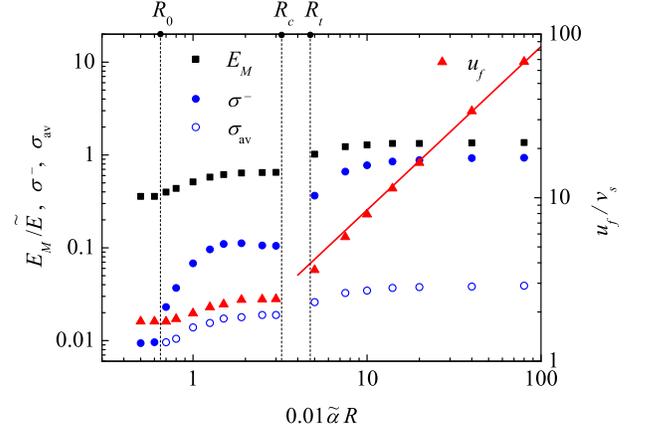}\\
\caption{Maximum field strength $E_M$, front velocity $u_f$, maximum $ \sigma ^ - $ and the average over the area $ \sigma_{av} $ concentrations of charge carriers behind  the front vs. parameter $R$ at $E_0=0.36\tilde{E}$.}\label{fig4}
\end{figure}

\begin{figure}[!t]\center
\includegraphics[width=230pt]{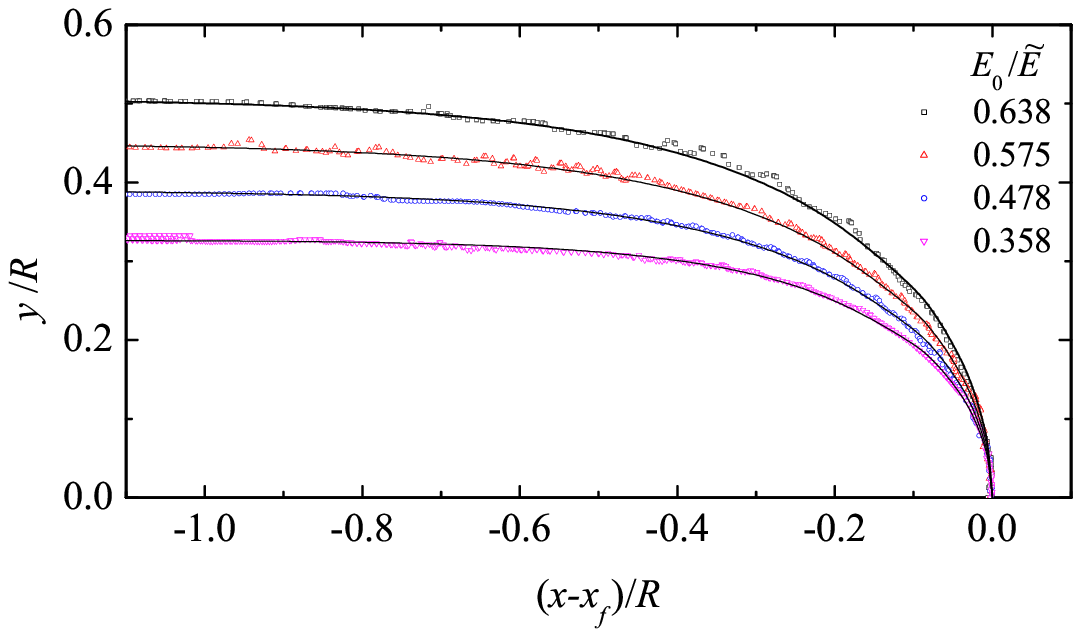}\\
\caption{Steady-front shapes of a hexagonal lattice of streamers with $\tilde{\alpha} R = 8000 $ and various external field strengths $ E_0 $. The symbols indicate the positions of the points behind the front, where the field strength on the results of the numerical simulation is equal to $0.001\tilde {E} $. Lines: approximation by the formula (\ref{eq1-1}).} \label{fig5}
\end{figure}

\begin{figure}[]\center
\includegraphics[width=230pt]{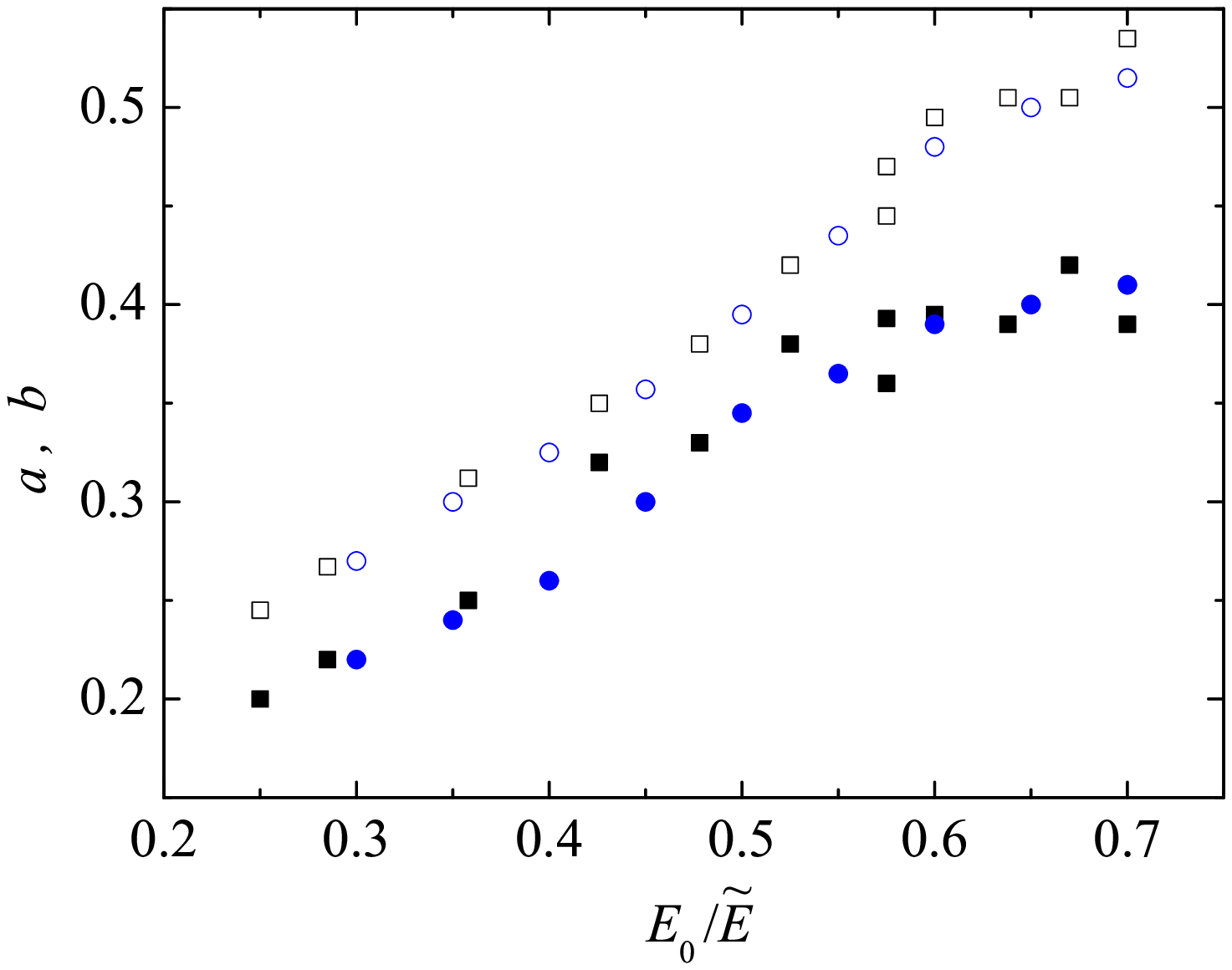}\\
\caption{The parameters $ a $ (filled symbols) and $ b $ (open symbols), which determine the steady-front shapes  of streamers array (see (\ref{eq1-1}), vs. the  external field strength $E_0$ with $\tilde{\alpha}R=8000$ (squares) and $\tilde{\alpha}R=4000$ (circles).}\label{fig6}
\end{figure}

\begin{figure}[!t]\center
\includegraphics[width=230pt]{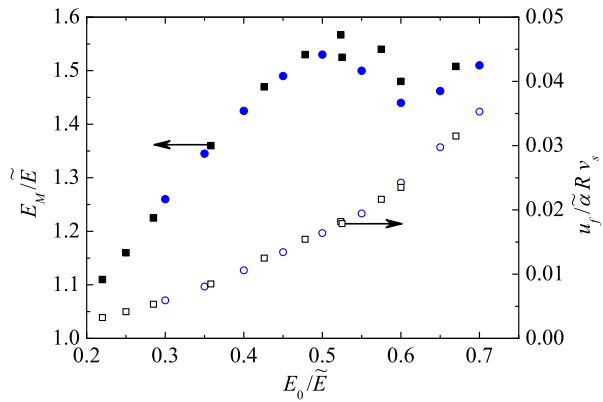}\\
\caption{Maximum field strength  $E_M$ and dimensionless time of flight $u_f/\tilde{\alpha R v_s}$ of the front distance $ R $ vs. the external field strength $E_0$  with $\tilde{\alpha}R=8000$ (squares) è $\tilde{\alpha}R=4000$ (circles).}\label{fig7}
\end{figure}

\begin{figure}[]\center
\includegraphics[width=230pt]{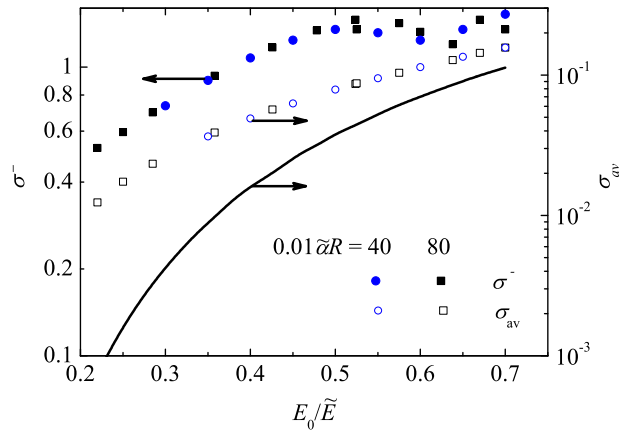}\\
\caption{Maximum $ \sigma ^ - $ and the average over the area $ \sigma_{av} $ concentrations of charge carriers behind  the front vs. the external field strength $E_0$  with $\tilde{\alpha}R=8000$ (squares) and $\tilde{\alpha}R=4000$ (circles). Line: calculation of $\sigma^-$ for a plane wave.}\label{fig8}
\end{figure}

\begin{figure}[!t]\center
\includegraphics[width=230pt]{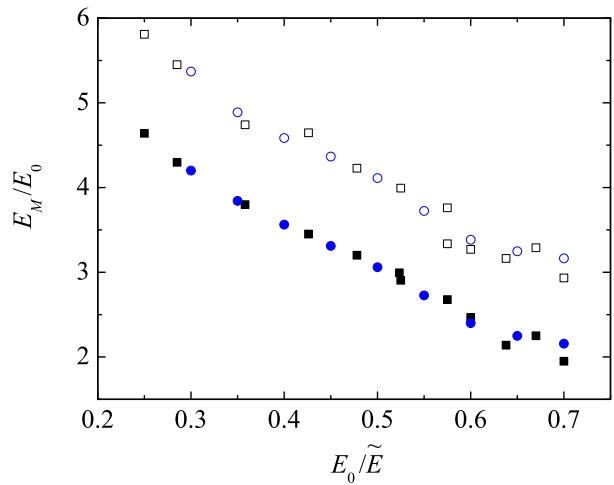}\\
\caption{Ratio $E_M/E_0$ vs. the external field strength $E_0$ with $\tilde{\alpha}R=8000$ (squares) è $\tilde{\alpha}R=4000$ (circles). Filled symbols: the simulation results, open symbols: calculation  formulas (\ref{eq3-15}) using the values of $a$ and $b$, shown in Fig. \ref{fig6}.}\label{fig9}
\end{figure}

\begin{figure}[]\center
\includegraphics[width=230pt]{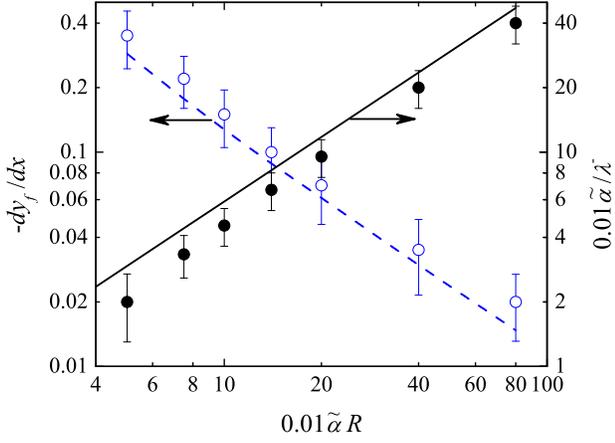}\\
\caption{The derivative of $ dy_f / dx $ in the region of linearity of function $ y_f (x) $ (open symbols) and the decay length  $ 1 / \lambda ^ - $ of  the field behind the front (filled symbols) vs. $ R $ with $E_0=0.36\tilde{E}$. Symbols are obtained by processing the simulation results, dashed line: calculation  formulas (\ref{eq3-8}) using the values of $ u_f $, shown in Fig. \ref{fig4}, solid line: calculation  formulas (\ref{eq3-6a}) using the values of $ b $, shown in Fig. \ref{fig6}.}\label{fig10}
\end{figure}

\begin{figure}[!t]\center
\includegraphics[width=230pt]{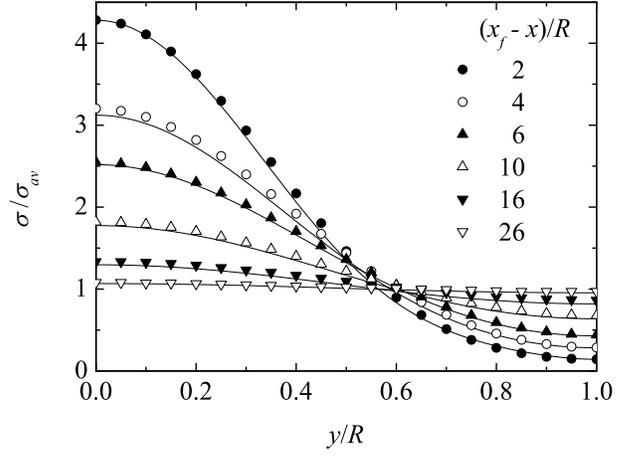}\\
\caption{The radial  distribution  of the charge carriers density $ \sigma $ at different distances from the front with $R= 100\tilde{\alpha}$ and $E_0=0.36\tilde{E}$. Line: the simulation results, symbols: the calculation formulas (\ref{eq3-10}) with $\sigma_{av}=0.0138$, $\sigma_{00}=0.059$ and $\sigma_{0R}=0.002$.}\label{fig11}
\end{figure}

\begin{figure}[!t]\center
\includegraphics[width=242pt]{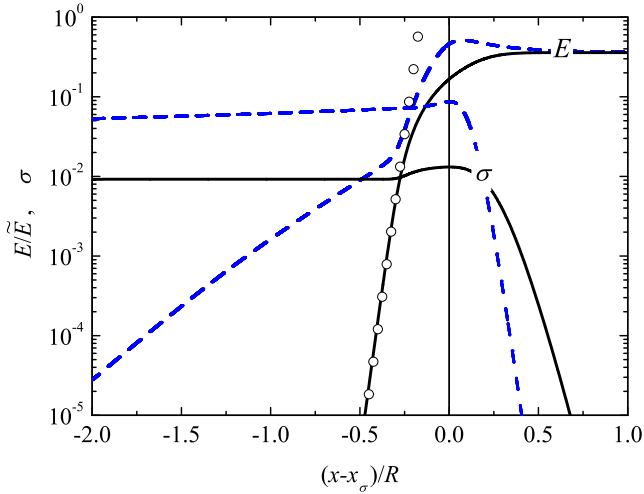}\\
\caption{Distributions of the field strength $ E $ and charge carriers concentration $ \sigma $   along the $ x $-axis near the fronts of the plane (solid lines) and corrugated (dashed lines) waves with $R= 100\tilde{\alpha}$ and $E_0=0.36\tilde{E}$. Symbols: the calculation formulas (\ref{eq3-14}) using the values of $\sigma^{-}$ and $u_f$, obtained in the simulation of a plane wave. Concentration $\sigma$ on the axis of $ x $ is maximal at $x=x_\sigma\lesssim x_f$.}\label{fig12}
\end{figure}

\begin{figure}[]\center
\includegraphics[width=242pt]{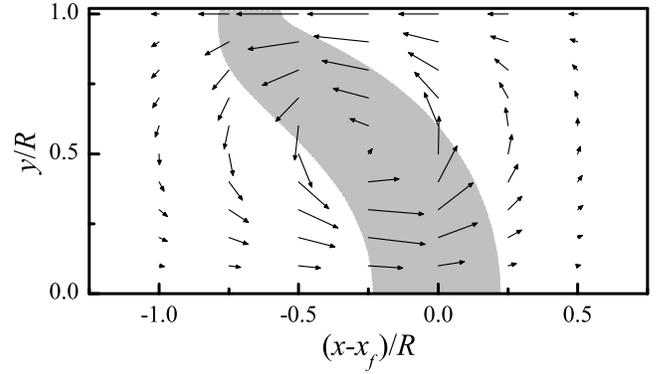}\\
\caption{The vector field of the current near the front of   corrugated ionization wave at $R= 100\tilde{\alpha}$ and $E_0=0.36\tilde{E}$. Length of the arrows is proportional to $yJ$, where $\bf{J}$ - the sum of the conduction and displacement current densities. The space charge density is more then $0.002\varepsilon\tilde{\alpha} \tilde{E} \approx 3,2$~mC/cm$^3$ inside the region shaded in gray.}\label{fig13}
\end{figure}

The calculations were performed for values $E_0=(0.2-0.7)\tilde{E}$ and $R=(20-8000)\tilde{\alpha}^{-1}$. The simulation results are presented in Figures \ref{fig1}-\ref{fig13}. They are very weakly dependent on the choice of the quantities $r_\sigma$ \cite{ASK7} and $\sigma_{th}=qn_{th}/\varepsilon \tilde{\alpha}\tilde{E}$ \cite{ASK6}; in the present study, we used the values $r_\sigma=2/\tilde{\alpha}$ and $\sigma_{th}=10^{-8}$.

As might be expected, avalanches and streamers develop independently from each other as described in \cite{ASK6}, while their length $ 2x_f \ll 2R $. At $ x_f = (2.5 \div 3 ) R $ an electrostatic interaction between them becomes significant (see Appendix). At first, it slows down the expansion of streamers (Fig. \ref{fig1}) and reduces the maximum field strength $ E_M $ at the front (Fig. \ref{fig2}). Further results of the interaction depend strongly on the values of the control parameters  $ E_0 $ and $ R $. It turned out that the plane of $ [1/E_0, R] $ splits into four areas shown in Fig. \ref {fig3}, in which the character of  streamers array evolution  is qualitatively different.

In regions 1 and 2 (i.e., at $ R> R_t (E_0) $ and $ R_0 (E_0) <R <R_c (E_0) $) expansion of streamers with time stops, their form and parameters ($ E_M $, $ \sigma ^ - $, $ u_f $) stop changing, that is, the entire array of streamers becomes a stationary ionization  wave, with more or less strongly curved (corrugated) front (see Fig. \ref{fig1}). For these regions  the tendency to decrease of  parameters $ E_M $, $ \sigma^ - $, $ u_f $  and the front curvature along  together with $ R $ is common (see Fig. \ref{fig4}).

For a description of these waves it is convenient to use the dimensionless coordinate system $ {\bf \hat {r}} = ({\bf r}-u_f x) / R $, moving with the front, and look for a solution in the form of

\begin{eqnarray}
\left|\begin{array}{c} \sigma(t,{\bf r})\\ \rho(t,{\bf r})\\ \varphi(t,{\bf r})\\{\bf E}(t,{\bf r})\end{array}\right| = \left|\begin{array}{c}
\hat{\sigma}({\bf \hat{r}})\\ (\tilde{\alpha}R)^{-1} \hat{\rho}({\bf \hat{r}})\\ \tilde{E}R\hat{\varphi}({\bf \hat{r}})\\ \tilde{E}{\bf F}({\bf \hat{r}})\end{array}\right|
\label{eq3-1}
\end{eqnarray}
The substitution of (\ref{eq3-1}) into (\ref{eq2-1}) - (\ref{eq2-3}) leads to a system of equations
\begin{eqnarray}
\frac{u_f}{R}\frac{\partial\hat{\sigma}}{\partial \hat{x}}+2(\hat{\sigma} v \alpha+ \frac{q g_t}{\varepsilon\tilde{\alpha}\tilde{E}})=\frac{1}{R^2}\left[\frac{1}{\tilde{\alpha}}\hat{\nabla}\mkern -4mu\cdot\mkern -3mu(\hat{\rho}{\bf v})-D\hat{\triangle}\,\hat{\sigma}\right],\label{eq3-2}\\
\frac{u_f}{R}\frac{\partial\hat{\rho}}{\partial \hat{x}}-\hat{\nabla}\mkern -4mu\cdot\mkern -3mu(\hat{\sigma}{\bf v})=-\frac{D}{\tilde{\alpha}R^3} \hat{\triangle}\,\hat{\rho},\qquad\qquad\label{eq3-3}\\
\hat{\nabla}\mkern -4mu\cdot\mkern -3mu {\bf F}=\hat{\rho}=
-\hat{\triangle}\,\hat{\varphi},\qquad\qquad\qquad
\label{eq3-4}
\end{eqnarray}
where differential operators $\hat{\nabla}$ and $\hat{\triangle}$  are with respect to $\hat{\mathbf{r}}$.

In region 1 the maximum field strength at the front $ E_M $ reaches a sufficiently large value (of the order of $\tilde{E}$ with our chosen values of semiconductor parameters) for  tunneling ionization to become noticeable. For very large $ R $ shape of the front $ y_f (x) $ and  parameters $ E_M $, $\sigma ^ - $ are independent on $ R $, the space charge density $ q \rho \propto 1/R $, the velocity  $ u_f \propto R $ and much higher than the maximum drift velocity $ v_s $ (see Fig. \ref{fig4}). These scaling laws are a direct consequence of the structure equations (\ref{eq3-2}) - (\ref{eq3-3}): their right-hand sides are negligible at $ R \rightarrow \infty $, and the dimensionless function $\hat{\sigma}({\bf \hat{r}}),  \hat{\rho}({\bf \hat{r}}), \hat{\varphi}({\bf \hat{r}}), {\bf F}({\bf \hat{r}})$ cease to depend on $ R $, if the  the front velocity $u_f\propto R$.

It is interesting to note that the solution (\ref {eq3-1}) is quite similar to the exponentially self-similar solution (4) of \cite {ASK7}, which describes evolution of a stand-alone streamer in a uniform field in infinite space. The only distinction is that exponentially  growing spatial scale in our case doesn't depend on time and is equal to $R$. Therefore rise-time of spatial scale, entering into the exponentially self-similar equations (11),(12) of \cite {ASK7}, is replaced  by time  for which the front moves ahead on distance  $R$ in equations (\ref {eq3-2}), (\ref {eq3-3})  without their right parts. The principle is that both types of self-similarity realize only for sufficiently large $ R $ and $ u_f $, when the terms with the factors $ R ^ {-2} $ or $ u_f ^ {-2} $ can be neglected. The physical meaning of this approach lies in the fact that for large $ R $ front velocity $ u_ {f} $ is a lot more than the average directed velocity of the charge carriers. In this limit, their transport is not directly involved in the variation of $\sigma $, which is caused exclusively by the ionization as described by (\ref{eq3-2}). The role of the drift is reduced to the formation of the space charge ( see equation (\ref{eq3-3})), that suppresses the field behind the front according to Poisson equation (\ref{eq3-4}), and determines the structure of a wave as a whole.

Dependencies $ y_f (x) $ for such distant streamers shown in Fig. \ref{fig5}, are well described by a function (\ref{eq1-1}) after replacement of $ L $ to $ R $. However, for the reasons mentioned in the introduction, the function (\ref{eq1-1}) should be considered only as one of the suitable approximations of the simulation results. Fitting parameters $ a, b $ included in it, depend on $ E_0 $ essentially (see Fig. \ref{fig6}), at that $ a \neq b $ and $ (a+b) \neq 1 $. Dependencies of other front parameters on $ E_0 $ are shown for this case in Fig. \ref{fig7}-\ref{fig9}.

As expected, the maximum field strength at the front $ E_M $ increases with $ E_0 $ (Figure \ref{fig7}); along with $ E_M $ dimensionless time $ u_f / \tilde{\alpha} R v_s $   for which the front moves ahead on distance  $R$ (Fig. \ref{fig7}) and the concentration of $ \sigma^-, \sigma_{av} $ (Fig. \ref{fig8}), all of which are independent on $ R $,  increase too. However, the ratio $ E_M/E_0 $ decreases (see Fig. \ref{fig9}). To explain this effects, it should be noted that with increasing $ (x_f - x) $ field $ E_n $ and, hence ionization rate decreases rapidly and becomes negligible at $x<x_i$, where  $E_n<E_i\equiv E_n(x_i)\sim 0.1\tilde{E}$. It is clear that the length  $ (x_f-x_i) $ of ionization region increases with $ E_M $ and $ E_0 $, so  radius $ bR $ of each streamer in the array must also increase. The parameter $ a $, characterizing the degree of "sharpness" of a streamer front  \footnote{The more $ a $, so the streamer front is sharper at a fixed value $ bR $.}, also increases with $ E_M $, but more slowly than $ b $. In other words, this means that   streamers have to become sharper, but thicker when $ E_0 $ rises. The results of numerical calculations of $ E_M $ for the simplest model of the streamers (see Appendix) show that the combined effect of these two factors should lead to falling dependence  $ E_M/E_0 $ on $ E_0 $, which is consistent with the simulation results (see Fig. \ref{fig9}) in qualitative terms. Some excess of the expected relations over  observable $ E_M/E_0 $ is due to the fact that the front of the streamer has a finite thickness $ \delta $, and the ratio of $ \delta / R $ increases with $ E_0 $ \cite {ASK10}.

Here we must mention an important result for practical applications: the average concentration  $ \sigma_{av} $ of electrons and holes behind a corrugated ionization wave is much more than behind the front of a plane wave (Fig. \ref{fig8}). The reason for this is that the reduction of the ionization area  $ b^2 $ times in a corrugated wave compared with a flat one is  compensated by an exponential increase of impact ionization rate. This effect is particularly large in a relatively weak external field, where the dependence of $\alpha(E)$ is very sharp.

With decreasing $ R $ it becomes apparent that the front resembles not  a cylinder with an oval tip, but a badminton shuttlecock (Fig. \ref{fig1}), and the slope of almost linear section of $ y_f (x) $ (shuttlecock feathers) decreases approximately inversely proportional to $ R $ (Fig. \ref{fig10}), and approximation (\ref{eq1-1}) is becoming less suitable for small $R$. To explain this effect, it should be noted that in the case of stationary wave the kinematic relation
\begin{equation}
\frac{dy_f}{dx}=-\frac{u_n}{\sqrt{u_f^2-u_n^2}}
\label{eq3-5}
\end{equation}
holds, where  $u_n=u_n[E_n(x)]$ \footnote{If the wave propagation is significantly affected by tunnel ionization, velocity $ u_n $ depends not only on  $ E_n $, but also on the whole field distribution along a force line  intersecting the front \cite{ASK7, ASK10}. This makes the analytical calculation of the whole front form quite so hopeless, but has no effect on the properties of the function $y_f(x)$ at $x<x_i$.} and  $y_f(x_f)=0$, $u_n(x_f)=u_f$ by definition. In the area of $ x <x_i $ the front itself ceases to be an ionization front, and it is just a thin layer of a space charge, moving mainly due to an electron (if  $ \rho <0 $ at the front) or holes (if $ \rho > 0 $ at the front) drift, so $u_n=v[E_n(x)]$. From this and the scaling laws $ u_f \propto R $, $ E_M = \tilde {E} F (0,0) $, it follows that the derivative of $ dy_f / dx $ tends to zero, and the streamers radius  tends to a constant value of $ b R $ at  $R\rightarrow \infty$ and $u_f/v_s\rightarrow \infty$. It is easy to show that in this case the field $ E_n $ behind the front decreases like
\begin{equation}
E_n(x)\approx E_i \exp[\lambda_E(x-x_i)],
\label{eq3-6}
\end{equation}
where
\begin{equation}
\lambda_E \approx \frac{1}{R}\left(\frac{1.85}{1-b}-1\right)
\label{eq3-6a}
\end{equation}
is a minimal positive root of the equation
\begin{equation}\nonumber
J_0(\lambda_E b R) Y_1(\lambda_E R)=  J_1(\lambda_E R) Y_0(\lambda_E b R),
\end{equation}
$J_\nu$ and $Y_\nu$ are Bessel functions of the first and second kind of order $\nu$. The substitution of (\ref{eq3-6}) in (\ref{eq3-5}) leads after integration to  asymptotic dependence
\begin{equation}
y_f(x)=b R-\frac{v}{\lambda_E \sqrt{u_f^2-v^2}}\ln\left[1+\frac{E_i}{E_s}e^{\lambda_E(x-x_i)}\right],
\label{eq3-7}
\end{equation}
which is valid under the condition that the second term is small in comparison with the first term. In semiconductors, it is usually $ E_i \gg E_s $, so there is an area where $E_s\ll E_n\leq E_i$, $v\approx v_s$ and
\begin{equation}
y_f(x)\approx y_f(x_i)-\frac{v_s}{\sqrt{u_f^2-v_s^2}}(x-x_i)
\label{eq3-8}
\end{equation}
according to the simulation results (see Fig. \ref{fig1}, \ref{fig10}). The formula (\ref{eq3-8}) should also be straight from (\ref{eq3-5}), and therefore, in contrast to (\ref{eq3-7}), is valid even if the ratio of $ u_f / v_s $ is not very large \footnote {It is interesting to note precise analogy between this ionization front and the shock waves produced by supersonic motion of bodies in gases.}. Consequently it is advisable to use an approximation
\begin{eqnarray}
y_f(x)=\frac{2}{\pi}b_\infty R\arccos\left[\exp\left(\frac{\pi}{2}\frac{x-x_f}{aR}\right)\right]-\nonumber\\
-\frac{v_s}{\sqrt{u_f^2-v_s^2}}(x-x_f),
\label{eq3-8a}
\end{eqnarray}
which is consistent with (\ref{eq3-8}) in the region  $(x-x_f)>2aR$ at an appropriate choice of $b_\infty$, coincides with (\ref{eq1-1}) at $v_s/u_f \rightarrow 0$ and well describes the shape of the front at all ratio $ v_s / u_f $ as long as $E_n>E_s$.

At more higher values of $ (x_f-x) $  the inequality $ E_n \ll E_s $ is satisfied, so it follows from  (\ref{eq3-7}), that function $ y_f (x) $ seeks to $ bR $  exponentially:
\begin{equation}
y_f(x)=bR-\frac{\mu_0 E_i}{\lambda_E u_f}e^{\lambda_E(x-x_i)},
\label{eq3-9}
\end{equation}
where $\mu_0=v_s/E_s$ is low-field mobility. In gases $E_s\gg E_i$ (that is $\mu=\mu_0$ in the topical range of fields), therefore the range of linearity of $ y_f (x) $ has to be absent, which is consistent with the results \cite{Luq1}. In a planar case $\lambda_E = \pi/2L(1-b)$, so that the formula (\ref{eq3-9})  correctly describes the asymptotic of function (\ref{eq1-1}) at $(x_f - x)\rightarrow \infty$, obtained as a result of an exact solution of the corresponding problem. This coincidence confirms the correctness of the above approximate method for determining the shape of the front at $x<x_i$.

However, it must be born in mind that due to the exponential decrease of the field strength $ E_n $ and the corresponding front surface charge density the front as such actually ceases to exist just where formula (\ref{eq3-9}) is formally applicable for semiconductors. Instead, the filament of quasi-neutral plasma arises and fills the entire area of $ y <R $, but this filament is highly non-uniform in the transverse direction. The main transport mechanism of the electrons and holes are becoming ambipolar diffusion, which results in a radial spreading of plasma, so that $\sigma\rightarrow\sigma_{av}$ at $x\rightarrow -\infty$ (Fig. \ref{fig11}). This process is well described by the formula

\begin{eqnarray}
\sigma(x,y)=\sigma_{av}+B\exp\left(\frac{x_0-x}{x_1}\right)J_0\left(a_{11}\frac{y}{R}\right)+\nonumber\\
+C\exp\left(\frac{x_0-x}{x_2}\right)J_0\left(a_{12}\frac{y}{R}\right),\qquad\quad \label{eq3-10}\\
B=\frac{(\sigma_{00}-\sigma_{av})J_0(a_{12})-\sigma_{0R}+\sigma_{av}}{J_0(a_{12})-J_0(a_{11})}, \quad \nonumber\\
C=\frac{(\sigma_{00}-\sigma_{av})J_0(a_{11})-\sigma_{0R}+\sigma_{av}}{J_0(a_{11})-J_0(a_{12})}, \quad \nonumber
\end{eqnarray}
where $x_k=u_f R^2/D a_{1k}^2$ and $a_{1 1}\approx 3,83$,  $a_{1 2}\approx 7,02$ are the first and second roots of equation $J_1 (x)=0$, $\sigma_{00}$ and $\sigma_{0R}$ is concentration in a plane $x=x_0 < x_i$ at $y=0$ and $y=R$ respectively, which, like the $ \sigma_ {av} $, are determined by processing the simulation results. The formula (\ref{eq3-10}) is obtained using the first two terms of the series (8.3) from Chapter VII of the book \cite{Cars-Jaeg} and is applicable provided that $ x_1 \gg D / u_f $, when the longitudinal diffusion is negligible. The formula (\ref{eq3-10}) also helps to explain the fact that the transverse spreading of  plasma is almost imperceptible in region 1, but it becomes a determinant at small $ x_k \propto u_f R ^ 2 $, that is in region 2.

In addition, the reduction of $ R $ also leads to two effects. Firstly, $ E_M $ decreases so that when $ R < R_t (E_0) $ a tunnel ionization, ceases to have significant influence on the evolution of streamers array  and at a relatively weak external field (at $ E_0 <0.4 \tilde {E} $ in our case) the stationary distribution of streamers becomes impossible due to front transverse instability each of them. Secondly, an even greater reduction of $ R $  suppresses the instability at last and at $ R < R_c (E_0) $ (in region 2 in Fig. \ref{fig3}) evolution of streamers array again completes by the appearance of a stationary corrugated wave, which is  propagating now not only due to drift and impact ionization, but also due to diffusion. The same mechanism determines the evolution of isolated streamers during a "diffusion stage"  \cite{ASK6}. It is not surprising that the  streamer front instability  comes (according to \cite{ASK6}) where its radius is of the order of $ R_c $. It is interesting to note that the front of  two-dimensional streamers array  in gases also becomes unstable \cite{Luq1}, when  distance $ 2L $ between them is more than $ 2R_c $ (see Fig. \ref{fig3}). The boundaries of the instability region (region 4 in Fig. \ref{fig3}) is also  determined by the inequalities $ R_t (E_0)> R> R_c (E_0) $, so it exists only at $ E_0 <0.4 \tilde {E} $. In strong external fields tunneling ionization does not "turn off" and suppresses the instability even at $ R <R_c (E_0) $.

With further decrease of $ R $ curvature of a corrugated front and parameters $ E_M $, $ \sigma ^ - $ are also reduced. Primary avalanches begin to overlap until the avalanche-to-streamer transition and form a periodically perturbed  stationary ionization front. The most long-wave harmonic of this perturbation is given by the boundary conditions of our problem and has a wave number $ k = \pi / R $. Perturbation amplitudes (in particular, the difference $ E_M-E_0 $) decrease monotonically with $ R $ as long as  becomes vanish at $ R = R_0 (E_0) $. At still lower $ R $ (in region 3)  perturbations generated by the primary avalanche decay, so that over time a plane  impact ionization wave arises. This result is consistent with the conclusions of linear theory of transverse instability of the impact ionization waves. For gases, such a theory was created by the authors of \cite{Derks}, which showed that the perturbations with $ k <k_0 \approx \lambda ^ * /4 $ had to increase with time, and the more shortwave perturbations had to decrease. This approval is also true for semiconductors \cite{ASK8} if you use the appropriate  to our case formula \cite{ASK3,ASK2}
\begin{equation}\label{eq3-11}
\lambda^{*}=\alpha_0 \sqrt{\frac{v_s}{D\alpha_0}-\frac{1}{2}+\sqrt{\frac{1}{4}+\frac{v_s}{D\alpha_0}}},
\end{equation}
where  $\alpha_0=\alpha(E_0)$. This means that at
\begin{equation}\label{eq3-12}
R<R_0\approx 4\pi/\lambda^{*}
\end{equation}
an array of avalanches should produce a plane impact ionization wave \footnote{A similar effect was observed earlier in the simulation of flame propagation in a tube \cite{Bychkov96}: reduction of its diameter to  critical value corresponding of this process suppressed transverse instability and the flame front remains flat.} in accordance with the results of modeling. This wave propagates with the same rate
\begin{equation}\label{eq3-13}
u^*=\frac{D}{2}\left(3+\sqrt{1+\frac{4v_s}{D\alpha_0}}\right)\lambda^*,
\end{equation}
for all $ R <R_0 $  \cite{ASK3,ASK2} only due to drift, diffusion and impact ionization, as the field strength at its front does not exceed $ E_0 $ and is insufficient for tunneling ionization. The formula (\ref{eq3-12}) describes well the dependence of the critical radius $ R_0 $ (and coinciding with it the critical width $ L_0 $ for an array of flat streamers) on $ E_0 $, obtained from the simulation (see Fig. \ref{fig3}).

Another feature of the corrugated ionization waves  was found, but not explained in the paper \cite{Luq1}. It consists in the fact that the field behind the front tends to zero (in contrast to the isolated streamers \cite {ASK6, ASK7}), but decays much more slowly than in the front of the plane waves. A typical example of such differences is presented in Fig. \ref{fig12}. The paradox of this effect is the following. In the case of stationary plane ionization waves, as it is easy to show, the field strength is attenuated by the law $\exp(\lambda^- x)$  with \footnote{This formula is a bipolar analog of the formula (5.63) of the work \cite{Lag}.}
\begin{equation}\label{eq3-14}
\lambda^-=\frac{u_f}{2D}\left(\sqrt{1+4\sigma^- v_s\frac{\tilde{\alpha}\tilde{E}D}{E_s u_f^2}}-1\right)
\end{equation}
in exact agreement with the simulation results (see Fig. \ref{fig12}). The calculation of this formula for a corrugated wave always gives a significantly (almost four times in the case corresponding to Fig. \ref{fig12}) larger  value of $ \lambda^-$, but the simulation results give significantly (approximately ten times) smaller value of $ \lambda^-$.

The reason for this discrepancy is a curvature of the front. In a stationary wave  conduction current is accurately compensated by displacement current, so that the total current density is zero everywhere \footnote{If the field ahead of the front of a stationary plane wave is not uniform (as in $p^+ -n-n^+$-junction) then total current density is not zero, but is constant everywhere \cite{ASK4} and the field strength behind the front approaching to a finite value exponentially with exponent $\lambda^- $, that is also determined by the formula (\ref{eq3-14}).}.  The situation is completely different in  corrugated waves. Near $x$-axis ahead of the front conduction current and the displacement current coincide with the direction of wave propagation, but displacement current is  opposite directed and dominates near the surface $ y = R $ (where the concentration $ \sigma $ is very small). As a result, a vortex of current is formed at the front.  The example of such vortex is shown in  Fig. \ref{fig13}. Near the surface $ y = R $ field strength, and with it the displacement current decay exponentially according to (\ref{eq3-6}) with exponent $ \lambda_E $. Obviously, the conduction current,  providing the appearance of ohmic field behind the front, decreases in the same way near the axis of $ x $. It is clear that this ohmic field should decrease approximately exponentially with the increment of $\lambda_- \sim \lambda_E$. This conclusion is confirmed by the simulation results given in Fig. \ref{fig10}.

In conclusion of this section we will note two more circumstances. Firstly, at given  $ E_0 $ and $ R $ a stationary corrugated ionization wave  is an attractor for a wide range of initial conditions: after sufficient time the same solution is obtained

- for a point initial perturbation,

- for an ellipsoidal initial perturbation with a transverse semi-axis of order of $R/2$,

- as a result of restructuring a stationary corrugated wave with  control parameters $\{F'_0,R\}$ after changing  $F'_0$ to $F_0$,

- as a result of growth of small transverse perturbations of the plane ionization, if $ R <\pi / k_M $, where $ k_M $ is a wave number of the most rapidly growing perturbations \cite{ASK8}.

A similar (but less common) result was obtained for the two-dimensional array of streamers in gases \cite{Luq1}.

Secondly, apart from the above basic principles of evolution of interacting streamers  more subtle effects  were found. In particular, at some values of control parameters $ E_0 $ and $ R $, small quasi-periodic variations of the quantities $ \sigma $ behind the front and the maximum field strength $E_M$ are observed, as well as small quasi-periodic deviations of the front shape  from the linear function (\ref{eq3-8}). A possible reason for such anomalies is the use of a too coarse finite element mesh, the minimum size of which is determined by the resources of our computers. However, we can not exclude the physical reality of the observed effects (see, for example \cite{Lag}).

\section{Conclusion}

In the present article the results of numerical modeling of evolution of the two-dimensional periodic array of identical streamers in a constant and uniform field  are stated for the first time. It turned out that the nature of the evolution of a streamers array in semiconductors was much more complex in comparison with gases in a framework of "minimal model" \cite{Luq1}. The evolution of the array is completed in different ways, depending on the control parameters of the problem - the external field $ E_0 $ and the distance $ 2R $ between the streamers. For classification of various scenarios of evolution a diagram of final state  of the streamers array, dividing plane of $ [1/E_0, R] $ for four qualitatively various regions, is constructed and  represented on Fig. \ref{fig3}. In regions 1 and 2 interaction between streamers leads over time to formation of two types of stationary ionization waves with the corrugated front, differing with ionization mechanisms. Specific characteristics of fronts of these waves, caused by features of processes of ionization and charge transport in semiconductors, are described in detail and explained on the basis of simple physical reasons. In region 3, at enough small $R$, the array of primary avalanches generates a planar impact ionization wave. In weak external fields between regions 1 and 2 there is region 4 in which stationary propagation  of ionization  waves is impossible because of development of transverse instability. This instability observed earlier at modeling of isolated streamers, can be called as \emph{local} because it destroys (or doesn't destroy) fronts of each streamer in the array.

However, besides it \emph{global} instability of  array of streamers (each of which is locally stable) is also possible. In fact, in this work we considered that the primary avalanches generating the array of streamers, started \textbf{at the same time} from nodes of a \textbf{ideal} planar hexagonal lattice. Meanwhile small deviations in time and/or in the provision of avalanches start  will lead to emergence "competition" between streamers of the non-ideal array: some of them will appear in "preferred position"  and will  develop quicker than others. Sooner or later such streamers will start reducing considerably a field strength in the vicinity and to suppress propagation of neighbors.  If  distance $ d $ between the electrodes is sufficiently large, only the earliest and/or most remoted from neighbors streamers will be able to overcome it, and  the average distance between  such leading streamers should be of the order of $3d$ (see Appendix). These reasons indicate the importance of  a global instability problem which will be analyzed in a separate paper.

The author is grateful to A.V. Gorbatyuk for helpful  discussions. This work was supported by RFBR (grant 13-08-00474).

\begin{figure}[]\center
\includegraphics[width=230pt]{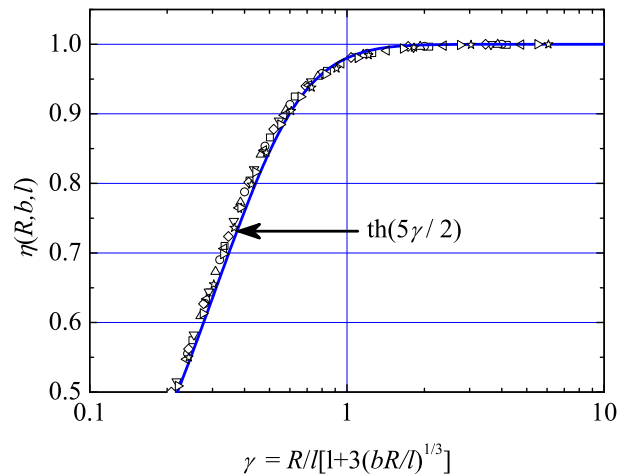}\\
\caption{Ratio of the maximum field strength at a single cylinder and an array of cylinders with hemispherical caps $\eta(R,b,l)$ vs. their length $2l$, radius $ bR $ and distance between them $2R$ with $bR/l=0.01-0.25$ and $R/l=0.4-50$.}\label{fig14}
\end{figure}

\begin{figure}[!t]\center
\includegraphics[width=230pt]{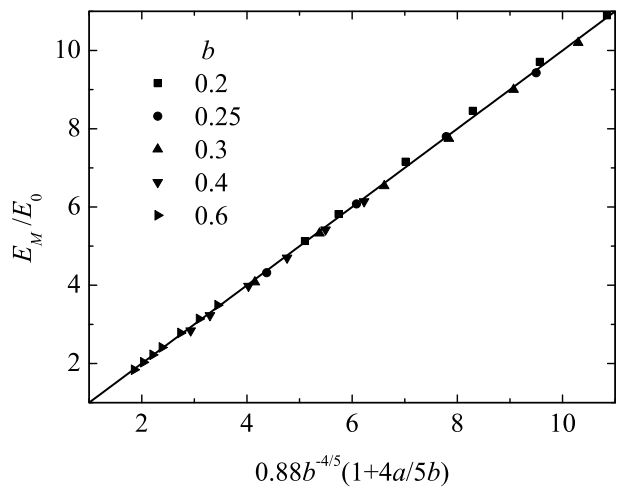}\\
\caption{Ratio $E_M/E_0$ for an array of cylinders with ellipsoidal tips vs. the dimensionless length of the longitudinal $ a_e = \pi a / 2 $ and transverse $ b $ semi-axes with $b=0.2-0.6$ è $a_e=0.2-1.2$.}\label{fig15}
\end{figure}

\section*{Appendix A. The electrostatic interaction between  metal cylinders with ellipsoidal tips.}

For a quick estimate of a maximum field strength  $ E_M $ of streamers in the array each of them can be represented in the form of a metal cylinder of length $ 2l $ with a radius $ bR $ and tips in the form of ellipsoids of rotation. To interpret the simulation results you need to know how the array parameters affect the two value.

The first of these - the ratio of $\eta(R,b,l)\equiv E_M(R,b,l)/E_M(\infty , b, l)$ - allows you to define the conditions under which the influence of the electrostatic interaction between  streamers on $ E_M $ becomes noticeable. These conditions depend little on the longitudinal axis of the ellipsoids $ a_e R $, therefore, when calculating $\eta$ it  would be possible to put $a_e =b$ for simplicity. The results of these calculations are shown in Fig. \ref{fig14}. As can be seen, for all relevant values of $ b $ and $ R / l $ they are well described by the function $\eta = \tanh(5\gamma/2)$, where $\gamma = R/l[1+3(bR/l)^{1/3}]$. Interaction becomes significant when $ \eta $ is noticeably different from the one, that is at $\gamma \lesssim 1$ or $R\lesssim (2,5\div 3)l$.

The second desired value - the ratio $E_M(b,a_e)/E_0$ at $l\gg R$ - allows you to evaluate $ E_M $ in a stationary propagating of streamer array (or, in other words, in a stationary corrugated ionization wave). In this case, the result depends strongly not only on  $ b $, but also on  $ a_e $. We used the value $a_e = \pi a/2$, since in this case the difference between the system "cylinder+ellipsoid"  and the surface of  revolution with forming a kind of (\ref{eq1-1}) is minimal and for all $ a, b $ do not exceed 5\%. The results of such calculations presented in Fig. \ref{fig15}, are well described by the dependence
\begin{equation}
E_M(a,b)=0.88 E_0 b^{-4/5}\left(1+4a/5b\right).
\label{eq3-15}
\end{equation}

\end{document}